\documentstyle[aps,prl,epsf,twocolumn,psfig]{revtex}
\def\beq{\begin{equation}}
\def\eeq{\end{equation}}
\def\beqa{\begin{eqnarray}}
\def\eeqa{\end{eqnarray}}
\def\l{\left}
\def\r{\right}
\def\bdi{\begin{displaymath}}
\def\edi{\end{displaymath}}
\def\ds{\displaystyle}

\begin{document}

\twocolumn[\hsize\textwidth\columnwidth\hsize\csname @twocolumnfalse\endcsname

\title{Heteropolymers in a Solvent at an Interface}

\author{ A.~Maritan, M.~P.~Riva, and A.~Trovato }

\address{International School for Advanced Studies (SISSA),\\
and Istituto Nazionale di Fisica della Materia,\\
Via Beirut 2-4, 34014 Trieste, Italy }

\address{The Abdus Salam International Center for Theoretical Physics,\\
Strada Costiera 11, 34100 Trieste (Italy) }

\maketitle

\begin{abstract}
Exact bounds are obtained for the quenched free energy of a polymer with 
random hydrophobicities in the presence of an interface separating a polar
from a non polar solvent.
The polymer may be ideal or have steric self-interactions. The bounds
allow to prove that a ``neutral'' random polymer is localized near the
interface at any temperature, whereas a ``non-neutral'' chain is shown to
undergo a delocalization transition at a finite temperature.
These results are valid for a quite general {\it a priori} probability
distribution for both independent and correlated  hydrophobic charges.
As a particular case we consider random 
AB-copolymers and confirm recent numerical studies.

\vspace{10pt}
\vspace{10pt}
\end{abstract}
]

The statistical behavior of heteropolymers has been intensively studied in
recent years \cite{Fol,GHLO,Sommer,Kan,SSEru,MoSte,var}. They model random
copolymers \cite{GHLO,Sommer} and to some extent protein folding \cite{Fol}.
For example, a chain composed by hydrophobic and hydrophilic (polar
or charged) components in a polar (aqueous) solvent evolves toward
conformations where the hydrophobic part are buried in order to avoid
water, wheras the polar part is mainly exposed to the solvent \cite{Cre}.  
This is what commonly happens to proteins and makes them soluble in
aqueous solutions. However other proteins (e.g. structural proteins) are 
almost insoluble under physiological conditions and prefer to form 
aggregates \cite{Cre}. Many of the proteins which are insoluble in water are
segregated into membranes which have  a lipid bilayer structure
\cite{Cre}. Membrane proteins have a biological importance at least as great
as those which are water soluble. Usually one distinguishes integral and non
integral membrane proteins according to whether the protein is most immersed
in the lipid bilayers or simply anchored to the membrane respectively (in the
latter case the protein is essentially water soluble) \cite{Cre}. An
analogous situation occurs for copolymers at interfaces separating two
immiscible fluids. If the solvents are selective (i.e. poor for one of the two
species and good for another), AB-copolymers are found to stabilize the 
interface \cite{BrDeGr}. Moreover, random copolymers have been claimed to
be more effective in carrying out this re-enforcement action \cite{DOKHJ}. 

\vskip 0.5cm

The simplest theoretical approach to the above problems has been proposed by
Garel {\it et al.} \cite{GHLO}. In the case of membrane proteins the finite
layer of lipidic environment is modeled as an infinite semi-space. Though a
quite rough approximation, this is the simplest attempt in capturing the
relevant features due to the competition  of different selective effects
\cite{Bon}.

We will study a lattice discretized version
of their model. The nodes of an $N$ links chain occupy the sites 
$\vec r_i =(x_{i1},\ldots, x_{id})$, $i=0,\ldots,N$ of a $d$-dimensional
hypercubic lattice.  A flat interface passing through the origin and
perpendicular to the $\vec u=(1,\ldots,1)$ direction separates a polar
(e.g. water), on the $\vec u\cdot\vec r>0$ side, from a nonpolar
(e.g. oil or air), on the $\vec u\cdot\vec r<0$ side, solvent.
The $i$-th monomer interacts with the solvent through its hydrophilic charge 
$q_i>0$ ($q_i<0$ means that it is hydrophobic) and  contributes to the 
energy with a term $-q_i{\rm sgn}(\vec u\cdot \vec r_i)$ \cite{note1}.
For simplicity we
can associate the charge to the link between adjacent positions on the chain
instead that to the single monomer.

The partition function of the model for a chain $W$ starting at position
 $\vec r$ is
\beq\label{1}
{\cal Z}(\vec r,\{q_i\})=\sum_{W:\vec r\to . }\exp {\l\{ \beta 
\sum_{i=1}^N q_i {\rm sgn} (\vec u\cdot\vec r_i)\r\}},
\eeq
where  $\beta^{-1}={k_BT}$.
If one sums over non-interacting (ideal) chains, then the lattice version
of the model introduced in Ref. \cite{GHLO} is recovered.
We will consider also the
more physical case where steric interaction among monomers does not allow for
multiple occupancy of lattice nodes, studied for a particular case in Ref.
\cite{Sommer}.

The free energy density of the system reads $f(\vec r,\beta)=-
\lim_{N\to\infty}\frac{1}{\beta N}
{\overline{\ln {\cal Z}(\vec r,\{q_i\})}}$, where ${\overline{\cdots}}$
denotes the quenched average over the distribution of the charges $\{q_i\}$.
In the following we will assume that  $\{q_i\}$  are independent random
variables having a Gaussian distribution of the form
\beq\label{P(q)}
P(q_i)=\frac{1}{\sqrt{2\pi \Delta^2}}\exp\l[-\frac{(q_i-q_0)^2}{2\Delta^2}\r].
\eeq

More general cases will be treated at the end. In particular considering 
charges not independently distributed is of interest for {\it
designed} sequences as  happens for real proteins. 

We will show that for a neutral chain ($q_0=0$) $f(0,\beta)<
f(\vec r,\beta)$ with $|\vec r|\geq N$, in the large $N$ limit, and for all
 $\beta$. The same holds also for $q_0\ne 0$ if $\beta>
\beta_{upper}(|q_0|,\Delta)$ with $\beta_{upper}\to 0$ if $\frac{|q_0|}
{\Delta}\to 0$. This implies that the chain is localized around the interface 
at any temperature if $q_0=0$, and at sufficently low temperature if  $q_0\ne 
0$. The proof is rigorous for the ideal chain, whereas for the self-avoiding
 case only a mild and well accepted hypothesis on the asymptotic behavior of
 the entropy is needed. When  $q_0\ne 0$ a rigorous lower bound on the
 free energy for both the ideal and the  self-avoiding chain allows to 
 determine a $\beta_{lower}(|q_0|,\Delta)$ below which the chain is 
delocalized. 
 
We will first consider the ideal chain case and then explain the modifications
necessary to extend the results to self-avoiding chains.

\vskip 0.5cm

{\em Ideal chain.} For clarity we derive the bounds in the $d=1$ case. The 
general case does not contain any further difficulty \cite{note2}.  
Let us first consider initial positions far from the interface in the
favorable solvent, $x\ge N$ if $q_0>0$ or $x\le -N$ if $q_0<0$. Under these
assumptions all chains remain in the same side, implying that ${\rm sgn}
 \l(x_i\r)=1$ (or 
${\rm sgn} \l(x_i\r)=-1$ respectively) for all $i$. 
Upon averaging over the charge distribution, we obtain the free energy density
of a walk in the favorable solvent:
\beq
f^*=-\frac{1}{\beta} \ln 2 - |q_0|.
\eeq

We give an upper bound to the free energy as follows. Consider only 
chains made up of blobs of $k$ steps, with $k$ even.
Bringing a blob in its globally favored side leads to an energy contribution
of the form $H_{j-{\rm th}\:{\rm blob}}
=-\l|\sum_{i=1}^{k} q_{k(j-1)+i}\r|$,
 so that 
\bdi
{\cal Z}(x=0,\{q_i\})\ge 
\l( C_k
\r)^{\frac{N}{k}} \exp\l\{ \beta \sum_{j=1}^{\frac{N}{k}}
          \l|\sum_{i=1}^k q_{k(j-1)+i}\r|\r\},
\edi
where $C_k$ is the number of chains starting and ending in the origin and
remaining in the same side. In the one-dimensional case it is easy to exactly
determine $C_k$.
It turns out  $C_k=\ds{\frac{1}{\frac{k}{2}+1}}
 \left( \begin{array}{c} k \\
\frac{k}{2}\end{array}\r)$, so that, by
using the Stirling's formula, the asymptotic result
$C_k\sim 2^k k^{-\frac{3}{2}}$ is found
(in $d$ dimensions $C_k\sim (2d)^k k^{-\frac{d+2}{2}}$ \cite{Hug}).
The upper bound on the free energy is then:
\beq
f(0,\beta)\le -\frac{1}{\beta} \frac{\ln C_k}{k} - \frac{1}{k} {\overline{
\l|\sum_{i=1}^k q_i \r|}},
\eeq
and, by using Eq. (\ref{P(q)}), we obtain: 
\beqa
\Delta f & = & f(0,\beta)-f^*\le h_{q_0}(k,\beta)\equiv\nonumber \\ & \equiv &
\frac{1}{\beta}\l[\ln2-\frac{\ln C_k}{k}\r] - |q_0|
G\l(\frac{\sqrt{k}|q_0|}{\sqrt{2}\Delta}\r),
\label{ub}
\eeqa
where the scaling function $G$ is given by
\beq
G(x) = \frac{1}{\sqrt{\pi}}\frac{1}{x} e^{-x^2}-[1-{\rm erf}(x)]\;.
\eeq
$G(x)$ is a positive decreasing monotonic function for positive arguments.

We consider separately the neutral and the $q_0\ne 0$ cases.
In the neutral case it turns out that the chain is always localized at the 
interface.  In fact, if $q_0=0$ we have
\beq\label{ubn}
\Delta f \le h_0\l(k,\beta\r)=
  \frac{1}{\beta}\l[ \ln 2- \frac{\ln C_k}{k}\r] -
  \sqrt{\frac{2}{\pi}} \Delta \frac{1}{\sqrt{k}}.
\eeq
It is easy to see that for any $\beta$ there exists a value $k(\beta)$
such that $h_0(k,\beta)<0$ for $k>k(\beta)$. For example at high temperature 
$k(\beta)\sim \l[\ln\l(\beta\Delta\r)\r]^2(\beta\Delta)^{-2}$.
This shows that at any temperature a 
neutral random chain is always adsorbed by the interface.

In the non-neutral case with $k=2$, one has $\Delta f<0$ if
\beq
\beta >\beta_{upper}=\frac{\ln 2} {|q_0| G(\frac{|q_0|}{\Delta})}.
\label{est}
\eeq
The limit $\lim_{k\to\infty}h_{q_0}(k,\beta)=0_+$
does not allow to deduce the existence 
of a negative minimum in $k$, so that the previous argument,
showing that the neutral chain is always localized, does not hold
for $q_0\ne0$.
Equation (\ref{est}) proves localization at sufficiently low temperatures.
For $|q_0|\ll \Delta$, it yields $\beta_{upper}=\frac{
\sqrt{\pi}\ln {2}}{\Delta}$, and in the opposite
 regime $\Delta\ll |q_0|$, 
$\beta_{upper}=2 \sqrt{\pi}\ln {2}e^{\frac{|q_0|^2}{\Delta^2}}
\frac{|q_0|^2} {\Delta^3}$.
In the limit $|q_0|/\Delta\ll1$ we can give a better estimate for
$\beta_{upper}$, such that $\beta_{upper}\to 0$ as $|q_0|\to0$,
by considering a larger blob
size $k=2x_0^2\l(\Delta/|q_0|\r)^2$, where $x_0\gg|q_0|/\Delta$ is fixed:
\beq
\beta_{upper}=\frac{3}{2x_0^2G(x_0)}
\ln\l(\sqrt{2}x_0\Delta/|q_0|\r)\frac{|q_0|}{\Delta^2}.
\label{best}
\eeq

We now look for a lower bound on $f$ for all chain initial positions.
If, for example, $q_0>0$, the preferred side is the right one.
Consider then the starting point at $x=N-k$ ($0<k\le N$) and let $g_E$,
with $E$ some subset of the last $k$ steps of the walk ($0<|E|\le k$ with
$|E|$ the number of elements in $E$), be
the number of walks having the steps belonging to $E$ in the unfavorable side.
Upon defining $G_k=\sum_E g_E$
($G_k<2^k$) one  has
\begin{eqnarray}\begin{array}{l}
{\cal Z}(x=N-k,\{q_i\})= \l(2^N-G_k\r)e^{\beta \sum_{i=1}^{N} q_i}+
\vspace{0.4cm}\\ \hspace{1.2cm}
+ e^{\beta \sum_{i=1}^{N} q_i}\ds{\sum_E g_E e^{ 
-2 \beta \sum_{i\in E} q_i}},
\end{array}
\end{eqnarray}
so that, by using the inequality $\ln {\overline {x}}\ge {\overline{ \ln x}}$
and averaging over the charge distribution, one obtains
\bdi
f(N-k)\ge f^* - \frac{1}{\beta N}\ln \l [ 1 + \sum_E g_E\frac {
a(\Delta,q_o,\beta)^{|E|}-1}{2^N}\r],
\edi
with $a(\Delta,q_o,\beta)=\exp\l[2\Delta^2\beta^2-2\beta q_0\r]$.
This equation and its analogous in the $q_0<0$ case show that
there is a delocalization temperature
\beq
\beta_{lower}= \frac{|q_0|}{\Delta^2}
\label{lowb}
\eeq
such that, if $\beta<\beta_{lower}$, $f(|N-k|)\ge f^*$ and the chain
delocalizes. This argument can be extended to the $d$-dimensional case, in
which $1\le G_k<(2d)^k$, ${\cal Z}_d^*(q_i)= (2d)^N e^{\beta\sum_{=1}^N q_i}$
and $f_d^*=-\frac{1}{\beta} \ln {2d} - |q_0|$.

The bounds we have proved above allow to conclude 
that there is a critical value $\beta_c$ 
such that for values of $\beta$ smaller than $\beta_c$ the chain is 
delocalized in the favorable solvent, while for larger values it is adsorbed 
by the interface, with the estimates $\beta_{lower}<\beta_c<\beta_{upper}$
(it is easy to verify that $\beta_{lower}<\beta_{upper}$).
The lower bound (\ref{lowb}) and the upper bound (\ref{best}), in the limit
$|q_0|/\Delta\ll1$, show the same behavior found by using both
an Imry-Ma type argument \cite{GHLO} and variational approaches 
\cite{SSEru,var}.

\vskip 0.5cm

{\em Self-avoiding chain.}
All the results shown for a random chain can be readily
generalized for a self-avoiding chain. Namely, a neutral chain is localized
at all temperatures, whereas a non-neutral chain undergoes a localization
transition at some critical temperature $\beta_c$.

The delocalization temperature $\beta_{lower}$ can be derived exactly in the
same way, since the division of walks into classes
according to the number of steps made in the unfavorable solvent does not
depend on the self-avoidance constraint.

 The upper bounds on the free energy,
which allow to prove chain localization, requires instead some refinements
with respect to the previous case. While the energy term is computed in the 
same way as before, the entropy term is different. Firstly,
 the connective constant
($\kappa=2d $ for a random walk in $d$ dimensions) is different. We recall 
that the existence of the connective constant,  
   $\kappa=\lim_{N\rightarrow\infty}\ln S_N\:/N$,
for self-avoiding walks (SAW) has been rigorously established \cite{Ham}
($S_N$ is the total number of $N$-steps SAW starting from the same site).
 The subleading correction of the form
$S_N \simeq \kappa^N N^{\gamma-1}$ is widely agreed upon, although not
rigorously proved \cite{Hug}.
Secondly we introduce the notion of {\it loop}, following
e.g. \cite{Whit}, and consider only walks made up
of $N/k$ blobs, each blob being a $k$-loop, in such a way that
different blobs can be embedded independently, as well as for a random chain.
A $N$-loop is a $N$-steps SAW, starting and ending on the interface, which
always remains in the same half-space, with the further condition
$x_{01}-x_{02}\le x_{i1}-x_{i2}<x_{N1}-x_{N2}\;\forall i$.
It has been proved \cite{Ham2} that the free energy density of loops is the
same as for SAW, $\kappa_l\equiv\lim_{N\rightarrow\infty}\ln L_N \:/N=\kappa$,
where $L_N$ is the number of $N$-loops. The subleading correction is
usually assumed in the same form as for the number of SAW:
\beq
L_N\simeq\kappa^N N^{\gamma_s-1}\; .
\label{ga}
\eeq

These considerations are sufficient to generalize the previous results
to the self-avoiding case, yielding the following bounds for the critical
temperature:
\beq
\frac{\ln{\kappa}}
{|q_0|G(\frac{\sqrt{2}|q_0|}{\Delta})}\le\beta_c\le\frac{|q_0|}{\Delta^2}\; ,
\label{bou}
\eeq
which do not depend on the assumption (\ref{ga}) and is therefore rigorous.
Again, in the limit $|q_0|/\Delta\ll1$ a better estimate $\beta_{upper}$ can
be derived by using Eq. (\ref{ga}):
\beq
\beta_{upper}=\frac{1-\gamma_s}{x_0^2G(x_0)}
\ln\l(\sqrt{2}x_0\Delta/|q_0|\r)\frac{|q_0|}{\Delta^2}.
\eeq

\vskip 0.5cm

{\em Generic probability distribution}.
Up to now we have considered the hydrophobic charges as independently
distributed  Gaussian
random variables. Actually, the results we have proved do not depend on
this assumption. We will briefly sketch this in a few cases \cite{next}.

The argument showing localization at any temperature for a neutral chain
holds true, both for random and self-avoiding chains, if
$\overline{\l|\sum_{i=1}^kq_i\r|}\simeq\sqrt{k}$ as $k\rightarrow\infty$.
The central limit theorem ensures this for independent random variables having
a generic probability distribution with finite variance and null mean.
In the non-neutral case, the existence of a delocalization transition
 can be proved e.g. for a bimodal distribution.
This corresponds to the  more realistic
case of two kinds of monomers, one hydrophilic and the other 
hydrophobic. We thus consider  the  generic bimodal distribution \cite{Sinai}:
\beq
P\l(q_i\r)=\alpha\delta\l(q_i-q_+\r)+\l(1-\alpha\r)\delta\l(q_i+q_-\r)\; ,
\label{bim}
\eeq
with  $q_+,q_->0$. The probability distribution (\ref{bim}) has three 
independent parameters, and fixing the average charge $q_0=\alpha\l(q_+
+q_-\r)-q_-$ and the variance
$\Delta=\sqrt{\alpha\l(1-\alpha\r)}\l(q_++q_-\r)$ we are left with one
free parameter.
It is interesting to report the delocalization temperature $\beta_{lower}$,
which provides a good estimate for the critical temperature in the previous
cases:
\bdi
\beta_{lower}^{bim}=\frac{\sqrt{\alpha\l(1-\alpha\r)}}{2\Delta}\l|
\ln\l[1+\frac{q_0}{\sqrt{\alpha\l(1-\alpha\r)}\Delta-\l(1-\alpha\r)q_0}\r]\r|
\edi

Notice that in the limit of nearly neutral chain ($|q_0|\ll\Delta$) we
get $\beta_c^d\simeq \frac{|q_0|}{2\Delta^2}$, which is the same function of
$|q_0|$ and $\Delta$ as in the Gaussian case,  suggesting the existence  of
a  universal behavior. In the limit of nearly homogeneous chain 
($\Delta\ll |q_0|$ which implies
$\alpha\simeq0$ or $\alpha\simeq1$) instead
 $\beta_{lower}^{bim}=\frac{1}{2\l(q_++q_-\r)}\l|
\ln\l[\frac{\alpha}{1-\alpha}\frac{q_+}{q_-}\r]\r|$ diverges logarithmically
in contrast with the Gaussian case. 

We consider now the case in which the hydrophobic charges $\l\{q_i\r\}$ are 
not independent random variables, but are Gaussianly distributed
with $\overline{q_i}=q_0\;\forall i$,
$\overline{q_iq_j}-\overline{q_i}\:\overline{q_j}=M^{-1}_{ij}$.
We assume $M^{-1}_{ii}=\Delta^2\;\forall i$, in analogy with the
non-correlated case, and also translational invariance along
the chain for the correlation matrix: $M_{ij}=b(\l|i-j\r|)$.
One can prove that if long range correlations decay exponentially or 
even algebraically the neutral chain is again localized at all temperatures. 
In fact, by  assuming an algebraic decay, 
$b\l(r\r)\simeq r^{-\eta}$, it turns out that
$\overline {\l|\sum_{i=1}^k q_i\r|}\simeq k^{\delta/2}$
with $\delta=\min\l(\eta,1\r)$.
 Only if correlations are so strong that they do not vanish along the chain 
($\eta=0$), the chain does not localize at all temperatures.

In the non neutral case the existence of the transition can be proved.
For example the estimate of the delocalization temperature is
\beq
\beta_{lower}^{corr}=\min_E\l\{\frac{\l|E\r|\l|q_0\r|}
{\sum_{i,j\in E}M^{-1}_{ij}}\r\}\; .
\eeq
If the charges
are positively correlated ($M^{-1}_{ij}>0$ for $i\ne j$), chain localization
is more favored than in the non-correlated case, whereas if charges are
anti-correlated ($M^{-1}_{ij}<0$ for $i\ne j$) it is less favored.

\vskip 0.5cm

{\em Asymmetric interface potentials.} Finally we extend our demonstrations
to the random AB-copolymers studied by Sommer {\it et al.} \cite{Sommer}. 
Their model corresponds to consider the following Hamiltonian:
\beq
{\cal H}=\sum_i |q_i| [\lambda \theta (q_i)\theta (-\vec u\cdot\vec r_i)+
\theta (-q_i) \theta (\vec u\cdot\vec r_i)],
\label{asymm}
\eeq
with the charges distributed according to Eq. (\ref{bim}) with
$\alpha=1/2$ and $q_+=q_-$ ($|\lambda-1|$ measures the potential asymmetry).
Such an AB-copolymer (Eq. (\ref{asymm})) is equivalent
to a non-neutral chain in symmetric potentials ($\lambda=1$), a case that we
have already discussed. We have proved the existence of a delocalization
transition for a neutral chain also in the Gaussian case.
For both distributions, the delocalization 
temperature shows the behavior $\beta_{lower}\simeq \frac{|\lambda-1|}{\Delta}$
in the limit of nearly symmetric potentials ($\lambda\simeq 1$), in agreement 
with the scaling law and the numerical results found in \cite{Sommer}.
On the contrary, in the highly
asymmetric cases (small and large $\lambda$)
different asymptotic behaviors for $\beta_{lower}$ occur \cite{next}.

\vskip 0.5cm

To conclude, in this Letter we have proved several exact results on random 
heteropolymers in the presence of an interface. Namely, a neutral chain is
localized at all temperatures, whereas a charged chain delocalizes at a 
finite temperature. The results are quite general and hold for ideal and
self-avoiding chains, Gaussian and bimodal distribution with independent and 
correlated charges. Furthermore, our lower bounds for the transition 
temperature confirm previous estimates.

\vskip 0.5cm

We would like to thank Jayanth Banavar, Cristian Micheletti and Flavio Seno
for useful discussion and ongoing collaboration,
and Enzo Orlandini for bringing references \cite{Sinai} to our attention.

\eject

\begin{thebibliography}{99}
\bibitem{note1}
{The case where the energy is $-\lambda_+ q_i$ and  $\lambda_- q_i$
 ($\lambda_{+,-}>0$) when $\vec u\cdot\vec r_i>0$ and $\vec u\cdot\vec
 r_i<0$ respectively
is readily reduced to the one treated here apart from an additive
 constant and a redefinition of the ``charges'' 
$ q'_i=\frac{\lambda_++\lambda_-}{2}q_i$.}
\bibitem{note2}
{ Notice that in the off-lattice model of Ref. \cite{GHLO} the entropy has 
only a harmonic term (Edwards functional) and thus the $d$-dimensional 
case reduces trivially to the $d=1$ case. In the lattice model considered here
this is no more true.}
\end{thebibliography}
\end{document}